\documentclass[prb,twocolumn,psfig,show pacs]{revtex4}
\usepackage{amsfonts}
\usepackage{amsmath}
\usepackage{graphicx}
\usepackage{bm}
\usepackage{amssymb}
\usepackage{times}
\usepackage{dcolumn}
\usepackage{cases}


\begin{document}

\title{Quantum Phase Transition, O(3) Universality Class and Phase Diagram
of Spin-1/2 Heisenberg Antiferromagnet on Distorted Honeycomb Lattice: A
Tensor Renormalization Group Study }
\author{Wei Li, Shou-Shu Gong, Yang Zhao, and Gang Su$^{{\ast }}$}
\affiliation{College of Physical Sciences, Graduate University of Chinese Academy of
Sciences, P. O. Box 4588, Beijing 100049, China}

\begin{abstract}
The spin-$1/2$ Heisenberg antiferromagnet on the distorted honeycomb (DHC)
lattice is studied by means of the tensor renormalization group method. It
is unveiled that the system has a quantum phase transition of second-order
between the gapped quantum dimer phase and a collinear N\'{e}el phase at the
critical point of coupling ratio $\alpha_{c} \simeq 0.54$, where the quantum
critical exponents $\nu \simeq 0.69(2)$ and $\gamma \simeq 1.363(8)$ are
obtained. The quantum criticality is found to fall into the $O(3)$
universality class. A ground-state phase diagram in the field-coupling ratio
plane is proposed, where the phases such as the dimer, semi-classical N\'{e}%
el, and polarized phases are identified. A link between the present spin
system to the boson Hubbard model on the DHC lattice is also discussed.
\end{abstract}

\maketitle

\section{Introduction}

Since the discovery of high temperature superconductors, the two-dimensional
(2D) Heisenberg models have received particular attention in the past
decades. Several numerical works show that the spin-1/2 isotropic Heisenberg
antiferromagnet (HAF) on a square lattice exhibits an AF long-range order
(LRO) in the ground state \cite{J D Regert,W H Zheng}, although a
mathematically rigorous proof still lacks now. Various methods (e.g. the
spin wave analysis \cite{M. E. Zhitomirsky}, different numerical techniques
\cite{Andreas}, etc.) were also applied to investigate the properties of
this model. When the bond anisotropy is introduced, the magnetic
order-disorder quantum phase transition (QPT) can be identified \cite{Subir
Sachdev,Singh,Sven,Matsumoto,Sandro Wenzel}. Another intriguing 2D bipartite
lattice---the honeycomb (HC) lattice has also been studied with various
methods, such as quantum Monte Carlo (QMC) \cite{J D Regert}, series
expansion \cite{Oitmaa}, spin wave \cite{W H Zheng2} and newly proposed
tensor renormalization group (TRG) method \cite{H. C. Jiang}. These
investigations show that owing to the lowest coordinates among 2D lattices,
the system is more affected by quantum fluctuations, giving rise to the
spontaneous magnetization per site of the spin-1/2 HAF model on this lattice
smaller than that on a square lattice.

Recently, people have obtained a number of magnetic materials with distorted
honeycomb (DHC) lattices, such as MnPS$_{3}$ and FePS$_{3}$ \cite{joy}, Cu$%
_{2/3}$V$_{1/3}$O$_{3}$ \cite{Kataev}, Na$_{3}$Cu$_{2}$SbO$_{6}$ \cite{Miura}%
, and Mn[C$_{10}$H$_{6}$(OH)(COO)]$_{2}\times $ 2H$_{2}$O \cite{Ivan Spremo}%
, where the magnetic ions (e.g. Cu$^{+2}$ and Mn$^{+2}$) form an HC lattice
with different nearest-neighbor (NN) bonds. The magnetic properties of these
materials have been investigated experimentally. However, the theoretical
studies on the HAF model on the DHC lattice are still sparse. There is a
recent work that explores the ground state properties of the Heisenberg
model on a DHC lattice by QMC calculations \cite{F.-J. Jiang}, and the
order-disorder QPT in zero magnetic field has been observed. Nevertheless,
the magnetic properties of the model on such a DHC lattice in nonzero
external fields are not yet seen in literature. Therefore, in order to
understand the experimental observations profoundly, it should pay more
theoretical attention on the HAF model on the DHC lattice.

\begin{figure}[tbp]
\includegraphics[angle=0,width=0.85\linewidth]{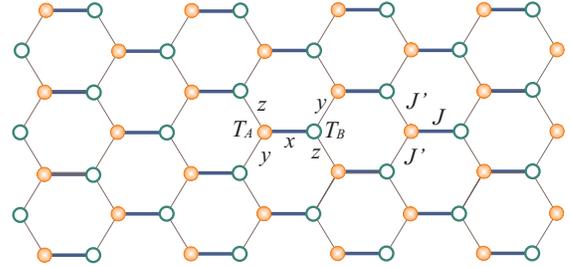}
\caption{(Color online) The distorted honeycomb lattice can be viewed as a
tensor network, where $T_{A}$ ($T_{B}$) are the tensors located on A (B)
sublattice (indicated by different symbols), and each tensor has three bond
indices labeled by $x$, $y$, and $z$.}
\label{Honeycomb lattice}
\end{figure}

In this paper, by means of the newly developed TRG method and a variational
analysis, we shall study the ground state properties of the spin-1/2 HAF
model on the DHC lattice with and without magnetic fields. On one hand, our
primary purpose is to understand the ground state properties of the model
under consideration in the presence of a magnetic field, and on the other
hand, we also want to test the accuracy of TRG methods in a more extensive
range rather than a spatially bond isotropic case by comparing our
calculated results with the previous studies on HC and DHC lattices using
other methods. Owing to the limitation of the present TRG scheme, a
finite-temperature calculation on the Heisenberg model is still not feasible
now, which makes it impossible to compare directly the calculated results
with the experiments. Our study shows that the spin-1/2 Heisenberg model on
a DHC lattice has a second-order QPT with respect to the bond coupling
ratio, that is determined to fall into the $O(3)$ universality class by
identifying the two quantum critical exponents. A phase diagram separating
the dimer, polarized, canted N\'{e}el and N\'{e}el phases is also proposed.
The magnetic properties in the presence of a magnetic field are obtained,
where some interesting behaviors are observed. In addition, the TRG method
has been verified through this present model to give a good agreement with
most of the previous studies, but it might overestimate slightly the
spontaneous sublattice magnetization per site on the HC lattice, implying
that the TRG method may still need more works for improvements.

The other parts of this paper is organized as follows. In Sec. II, the model
Hamiltonian and the TRG method will be introduced. In Sec. III, the magnetic
properties of the model under interest in the absence of a magnetic field
will be reexamined with the TRG algorithm. In Sec. IV, the ground state
properties in the presence of a magnetic field will be explored, and a phase
diagram will be proposed. In Sec. V, by invoking a boson mapping, the
present model is mapped onto the 2D boson Hubbard model, whose properties
will be briefly discussed. Finally, a conclusion will be given.

\section{Model Hamiltonian and numerical method}

The system under interest is schematically depicted in Fig. 1. The
Hamiltonian reads
\begin{equation}
H=J\sum_{\langle i,j\rangle _{x}}\mathbf{S}_{i}\cdot \mathbf{S}%
_{j}+J^{\prime }\sum_{\langle i,j\rangle _{yz}}\mathbf{S}_{i}\cdot \mathbf{S}%
_{j}-h\sum_{i}S_{i}^{z}-h_{s}\sum_{i}\epsilon _{i}S_{i}^{z},
\label{Hamiltonian}
\end{equation}
where $\mathbf{S}_{i}$ denotes the spin-1/2 operator at site $i$, $\langle
i,j\rangle _{x}$ labels the NN spins along the rungs ($x$ bond), $\langle
i,j\rangle _{yz}$ means the NN spins along the zigzag directions ($y$ and $z$
bonds), $J$ is the interaction on the $x$ bond, $J^{\prime }$ is the
coupling on the $y$ or $z$ bond, $h$ and $h_{s}$ stand for the uniform and
staggered magnetic fields, respectively, and $\epsilon _{i}=+1$ when $i$ on $%
A$ sublattice and $-1$ on $B$ sublattice. We introduce for convenience a
bond coupling ratio $\alpha =J^{\prime }/J$, and take $J$ as an energy scale.

To explore the ground state properties of the present system we shall
utilize the TRG method. This numerical algorithm was first introduced to
calculate the thermodynamic properties of the 2D classical models \cite{M.
Levin}, and then generalized to obtain the expectation values of observables
in the quantum state with the tensor product wave functions (e.g. Ref.%
\onlinecite{H. C. Jiang}) on bipartite lattices given by
\begin{equation}
|\Psi \rangle =\sum_{\{x_{i},y_{i},z_{i}=1\}}^{D}\prod_{i\in A,j\in
B}(T_{A})_{x_{i},y_{i},z_{i}}^{m_{i}} (T_{B})_{x_{j},y_{j},z_{j}}^{m_{j}}
|m_i m_j\rangle,  \label{TPS}
\end{equation}
where $T_{A}$ ($T_{B}$) represents the tensor located on $A$ ($B$)
sublattice, over which the indices $i$ and $j$ run, and the summation over
all bond indices $x,y,z$ is from 1 to the bond dimension $D$. According to
the TRG algorithm, the ground state wave function and energy can be directly
obtained by using trial wave functions of tensor product form \cite{Z. C. Gu}%
. This variational scheme, however, is not so efficient that makes the
achievable bond dimension $D$ not larger than 3 in general due to huge
variational parameter space. Recently, it was improved by combining the
infinite time-evolving block decimation (iTEBD) \cite{G. Vidal} and TRG
method to determine the ground state and to get the expectation values of
local observables \cite{H. C. Jiang}. This alternative algorithm appears to
be accurate and efficient, in which the available bond dimension $D$ can
reach as large as 8, and the calculated results agree well with those
obtained by other methods. It has been applied to study the spin flop
transition of spin-1/2 $XXZ$ model on a square lattice \cite{Pochung Chen}.
In the following, we shall adopt this novel scheme to calculate the physical
quantities of the spin-1/2 HAF system on the DHC lattice.

\begin{figure}[tbp]
\includegraphics[angle=0,width=1.1\linewidth]{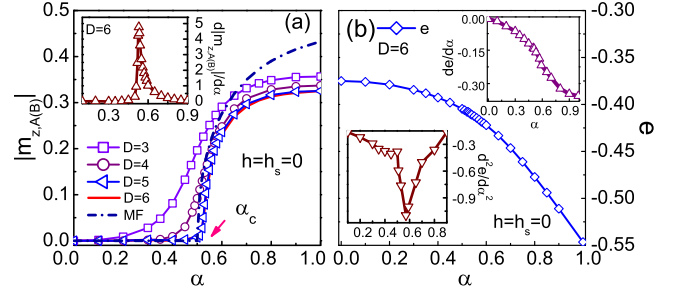}
\caption{(Color online) (a) The spontaneous sublattice magnetization per
site $|m_{z,A(B)}|$ as a function of the bond ratio $\protect\alpha $ for
different bond dimensions $D=3,4,5,6$. The dashed dot line represents the
mean-field result [Eq. (\protect\ref{Mvar})]. The inset gives the derivative
of $|m_{z,A(B)}|$ as a function of $\protect\alpha $, where a discontinuity
at $\protect\alpha _{c}\simeq 0.54$ is observed. (b) The ground state energy
per site $e$ and its derivatives (insets: first- and second-order) versus $%
\protect\alpha $, where the discontinuity in $\partial ^{2}e/\partial
\protect\alpha ^{2}$ against $\protect\alpha $ indicates a QPT of
second-order. }
\label{Order-Dis Tran}
\end{figure}

In our practical calculations, during the iterative projections by evolution
operator ($e^{-\tau H}$) along the imaginary time $\tau $ axis, we first
start with a step $\delta \tau =10^{-3}$, and then diminish it gradually to $%
\delta \tau =10^{-5}$. The total number of iterations is taken as about $%
10^{5}\sim 10^{6}$, where $D=5$ or $6$ is generally chosen. The convergence
is always checked, as shown in Fig. \ref{Order-Dis Tran}(a) for different
bond dimensions $D$.

\begin{figure}[tbp]
\includegraphics[angle=0,width=1.1\linewidth]{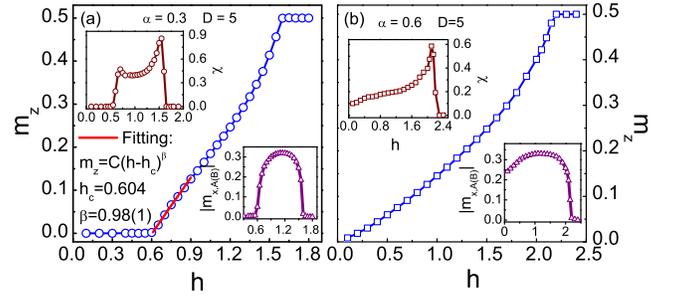}
\caption{(Color online) The magnetization per site as a function of magnetic
field $h$. The insets show the susceptibility $\protect\chi $ (upper panel)
and the transverse component of sublattice magnetization per site $%
|m_{x,A(B)}|$ (lower panel) as functions of $h$. (a) $\protect\alpha =0.3$;
(b) $\protect\alpha =0.6$, where $D=5$ and $h_{s}=0$ for both. The fitting
curves in (a) shows a nearly linear behavior of $m_z$, with $\protect\beta %
\simeq 0.98(1)$. The number in the parenthesis denotes numerical
fitting error hereafter.} \label{Magnetization}
\end{figure}

\section{Magnetic order-disorder transition}

Let us first consider the case in absence of a magnetic field ($h=0$ and $%
h_{s}=0$). When $\alpha =0$, the spins are coupled only by $J$ along $x$
bonds, and the ground state is $|\Psi _{g}\rangle = \prod_{i\in A}\frac{1}{%
\sqrt{2}}[|\uparrow _{i}\downarrow _{i+x}\rangle -|\downarrow _{i}\uparrow
_{i+x}\rangle ]$ with energy $-0.75J$ per bond, which is usually termed as a
dimer state. This disordered ground state is protected by a finite spin
singlet-triplet gap. When $J^{\prime }$ is set in, but $\alpha $ is still
small, one may conceive that the system may retain in the dimer phase\cite%
{Subir Sachdev}. This is confirmed in Fig. \ref{Order-Dis Tran}(a), where
the spontaneous sublattice magnetization per site, $m_{z,A(B)}=\frac{1}{%
N_{A(B)}}\sum_{i\in A(B)}\langle S_{i}^{z}\rangle $ with $N_{A(B)}$ the
total number of sublattice A(B) sites, as a function of $\alpha $ is
presented for $D=3,4,5,6 $. It is seen that there exists a critical ratio $%
\alpha _{c}$ below which $m_{z,A(B)}$ vanishes, showing that the ground
state for $\alpha <\alpha _{c}$ is disordered and dominated by quantum
fluctuations. For $\alpha >\alpha _{c}$, the ground state of the system has
an AFLRO owing to a spontaneous $SU(2)$ symmetry breaking. To determine the
value of $\alpha _{c}$ with accuracy, we have calculated the derivative of $%
|m_{z,A(B)}|$ with respect to $\alpha $ for $D=6$, and found a discontinuity
at $\alpha =\alpha _{c}$, where $\alpha _{c}$ can be readily determined as $%
0.54$.

To confirm if the transition occurring at $\alpha _{c}$ is a QPT, we have
also studied the ground state energy per site $e$ as a function of $\alpha $.
The results are presented in Fig. \ref{Order-Dis Tran}(b). It may be observed
that with increasing $\alpha $, both $e$ and its first derivative $\partial
e/\partial \alpha $ versus $\alpha $ decrease continuously, but the second
derivative $\partial ^{2}e/\partial \alpha ^{2}$ against $\alpha $ shows a
discontinuity at $\alpha _{c}\simeq 0.54$, as seen from the inset of Fig. %
\ref{Order-Dis Tran}(b). This feature characterizes a typical QPT of
second-order.

\begin{figure}[tbp]
\includegraphics[angle=0,width=0.8\linewidth]{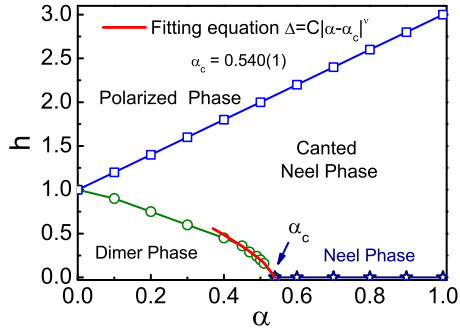}
\caption{(Color online) The ground state phase diagram of the spin-1/2 HAF
system on the DHC lattice in $\protect\alpha $-$h$ plane, where three phases
(dimer, semiclassical N\'{e}el and polarized) are identified, and the
collinear N\'{e}el phase marked by the star line inhabits exactly on the $%
\protect\alpha $ axis ($h=0$). The fitting curves reveal the critical
behavior in the region $\protect\alpha \in \lbrack 0.45,0.51]$.}
\label{Phase diagram}
\end{figure}

It should be noted that, as shown in Fig. \ref{Order-Dis Tran}(a), we get $%
|m_{z,A(B)}|\simeq 0.32$ \emph{($D=6$)} for $\alpha =1$, which appears to overestimate the
spontaneous sublattice magnetization per site on the HC lattice in
comparison to the recent stochastic series expansion QMC result 0.2681(8)
(Ref. \onlinecite{Low}), the "world line" QMC result 0.22 (Ref. %
\onlinecite{J D Regert}), series expansion 0.27 (Ref. \onlinecite{Oitmaa}),
and the spin-wave result 0.24 (Ref. \onlinecite{W H Zheng2}). Such a
discrepancy on $m_{z,A(B)}$ has also been noted in Ref. \onlinecite{Zhao},
where they reported the sublattice magnetization per site on the HC lattice
to be 0.3098 for $D=16$ by the TRG calculations. This slight discrepancy on
the sublattice magnetization per site may come from the underestimation of
quantum fluctuations in the absence of a magnetic field in the assumption of
the tensor product state employed in the TRG method, because the isotropic
system at $\alpha =1$ is gapless and has long-range correlations, where the
quantum fluctuations may be strong. However, the ground state energy per
site we obtained (-0.5465) for $\alpha =1$ \emph{($D=6$)} is quite consistent with those
obtained for the HC lattice by other methods, e.g. the QMC result -0.5450
(Ref. \onlinecite{J D Regert}), series expansion -0.5443 (Ref. %
\onlinecite{Oitmaa}), and the spin-wave result -0.5489 (Ref. \onlinecite{W H
Zheng2}). Therefore, as the critical point determined by the spontaneous
sublattice magnetization [Fig. \ref{Order-Dis Tran} (a)] coincides with that
obtained from the singularity of the ground state energy [Fig. \ref%
{Order-Dis Tran} (b)], it shows that the critical point is
determined with rather assurance. In the presence of a magnetic
field, since the quantum fluctuations are much suppressed, the
results given by the TRG method should be reliable. This can also be
validated in the following studies, including the attained linear
behavior of the magnetization curves immediately above the critical
magnetic field $h_{c}$ [Fig. \ref{Magnetization} (a)] and the
verified relation $H_{sat}=2S(2\alpha +1)$ of the saturation line
that separates the canted N\'{e}el phase and polarized phase in Fig.
\ref{Phase diagram}, which is also consistent with that derived from
the classical energy of spin wave analysis by a variational scheme
\cite{M. E. Zhitomirsky,Ivan Spremo}. Another fact is that when the
critical exponents of $\nu$ and $\gamma$ are determined (in Sec.
IV), the critical point is approached from the dimer phase, and
hence it is independent of the magnitude of spontaneous sublattice
magnetization in the N\'{e}el phase. The obtained $\nu$ and $\gamma$
agree well with previous calculations and theoretical predictions,
showing again that TRG method is fairly feasible for the present
case.

In order to examine our numerical results, we perform a mean-field treatment
in terms of a simple variational trial wave function
\begin{equation}
|\Psi _{var}\rangle =\prod_{i\in A}\frac{1}{\sqrt{1+t^{2}}}[|\uparrow
_{i}\downarrow _{i+x}\rangle -t|\downarrow _{i}\uparrow _{i+x}\rangle ],
\label{trial psi}
\end{equation}
that was applied to describe both disordered and ordered phases on a
dimerized square lattice \cite{Sven} and in a bilayer system \cite{C. Gros},
where the lattice site $i$ and $i+x$ are NN sites along $x$ bonds, and $t$
is a variational parameter and interpolates between a singlet collection ($%
t=1$) and a classical N\'{e}el state ($t=0$). Substituting Eq. (\ref%
{Hamiltonian}) (with $h,h_{s}=0$) and Eq. (\ref{trial psi}) into $%
E_{var}=\langle \Psi _{var}|H|\Psi _{var}\rangle $ and minimizing it with
respect to $t$, we obtain an upper bound for the ground state energy as
\begin{numcases}{E_{var}/NJ=}
-3/8, & for $\alpha \leq 0.5$ \nonumber \\
-\frac{1}{16}(1/\alpha+4 \alpha+2), & for $\alpha > 0.5$ \label{Evar}
\end{numcases}where $N=N_{A}+N_{B}$ is the total number of lattice sites.
Obviously, $E_{var}$ is singular at $\alpha _{c,var}=0.5$, showing $\alpha
_{c,var}$, that is close to $\alpha _{c}=0.54$, may be a transition point.
The variational sublattice magnetization per site $m_{var}=(1/N_{A})\sum_{i%
\in A}\langle \Psi _{var}|S_{i}^{z}|\Psi _{var}\rangle $ can be obtained by
\begin{numcases}{m_{var}=}
0, & for $\alpha \leq 0.5$\nonumber \\
\frac{1}{4}\sqrt{-1/\alpha^2+4}, & for $\alpha > 0.5$ \label{Mvar}
\end{numcases}which indicates that the derivative of
$m_{var}$ is discontinuous at $\alpha = 0.5 $, suggesting a
discorder-order phase transition at $\alpha _{c,var}$.
A comparison of $m_{var}$ to the TRG result is given in Fig. \ref{Order-Dis
Tran}(a), where $m_{var}$ shows a behavior similar to the TRG results.

\begin{figure}[tbp]
\includegraphics[angle=0,width=0.8\linewidth]{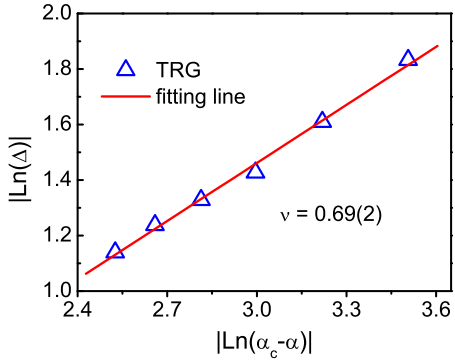}
\caption{(Color online) The log-log plot of spin gap $\Delta$ versus $%
\protect\alpha$ in the critical region $\protect\alpha \in \lbrack
0.45,0.51] $, where the linear fit gives the critical exponent $\protect\nu %
\simeq 0.69(2)$.}
\label{log-log-delta}
\end{figure}

\section{Ground-state phase diagram and critical exponents in presence of a
magnetic field}

Now we turn the uniform magnetic field $h$ on. The magnetization curves for
different $\alpha $ are given in Fig. \ref{Magnetization}. It is clear that
for $\alpha <\alpha _{c}$, there exists a magnetization plateau with $%
m_{z}=0 $, where $m_{z}=(1/N)\sum_{i}\langle S_{i}^{z}\rangle $ is the
magnetization per site, implying the existence of a finite spin gap, as
shown in Fig. \ref{Magnetization}(a) for $\alpha =0.3$. At $\alpha =0$, such
a gap is nothing but the spin singlet-triplet gap which equals $J$. For $%
0<\alpha \leq \alpha _{c}$, the spin gap will decrease and vanish eventually
at $\alpha =\alpha _{c}$. For a given $\alpha <\alpha _{c}$, the spin gap
closes at a critical field $h_{c}$. For $h\gtrsim h_{c}$, we find that the
magnetization depends almost linearly on the magnetic field, behaving $%
m_{z}\sim |h-h_{c}|^{\beta } $ with $h_{c}=0.604$, $\beta =0.98(1)$ for $%
\alpha =0.3$ and $D=5$. Our result is consistent with the theoretical
prediction $\beta =1.0$ in 2D and higher quantum spin systems \cite{Subir
Sachdev}. Nonetheless, it is in sharp contrast to the cases of gapped
one-dimensional Heisenberg spin systems where it is observed a square root
dependence of $m_{z}\sim |h-h_{c}|^{1/2}$ that characterizes the
commensurate-incommensurate phase transition \cite{R. Chitra}. For other $%
\alpha <\alpha _{c}$, such an almost linear dependence was also noted. The
susceptibility, $\chi =\partial m_{z}/\partial h$, as a function of the
magnetic field $h$ is shown in the upper inset of Fig. \ref{Magnetization}
(a). Two discontinuous points in $\chi $ versus $h$ are seen, namely one at
the point $h_{c}$ where the spin gap closes, and the other at the saturation
field. We have also explored the transverse component of sublattice
magnetization per site, $m_{x,A}=\frac{1}{N_{A}}\sum_{i\in A}\langle
S_{i}^{x}\rangle $, against the magnetic field $h$, as given in the lower
inset of Fig. \ref{Magnetization}(a). For $\alpha <\alpha _{c}$, $m_{x,A}$
vanishes for $h\leq h_{c}$, while it increases sharply when $h$ exceeds $%
h_{c}$, and after reaching a round peak it declines steeply and vanishes at
the saturation field. This observation shows that, with increasing the
field, there exists a transition from the disordered dimer phase to a canted
N\'{e}el phase (spin flop phase) \cite{M. E. Zhitomirsky,Andreas,Pochung
Chen} that is characterized by nonzero values of both $m_{z}$ and $m_{x,A}$,
where the spins align antiferromagnetically within the $xy$ plane and
develop a uniform $z$ component along the field, thus canting out of the
plane. For $\alpha >\alpha _{c}$, the magnetization curves behave
differently from those with $\alpha <\alpha _{c}$, as presented in Fig. \ref%
{Magnetization} (b) for $\alpha =0.8$ as an example. With increasing the
magnetic field, the magnetization $m_{z}$ increases monotonously till the
saturation, while the transverse component of magnetization $m_{x,A}$ first
increases slowly, then drops sharply and vanishes eventually at the
saturation field. The susceptibility $\chi $ increases slowly with
increasing the field, and then decreases steeply to zero at the saturation
field.

By summarizing the above observations, a ground-state phase diagram of the
system in the $\alpha -h$ plane can be drawn, as presented in Fig. \ref%
{Phase diagram}, where the phase boundaries are determined by the transition
points in Figs. \ref{Order-Dis Tran} and \ref{Magnetization}. One may see
that there are three phases, namely, the dimer phase, the semi-classical N%
\'{e}el phase (including the canted and collinear N\'{e}el states)
and the polarized phase. At $\alpha =\alpha _{c}$, there is a QPT
from the disordered dimer phase to the collinear N\'{e}el phase.
Note that the lower phase boundary between the dimer and canted
N\'{e}el phases is determined by observing the spin gap $\Delta $
that is obtained by calculating the width
of zero magnetization plateau for various $\alpha $ as presented in Fig. \ref%
{Magnetization} (a). The critical behavior of the spin gap in the dimer
phase in the vicinity of $\alpha _{c}$ is fitted by the least squares method
with $\Delta \sim (\alpha -\alpha _{c})^{\nu }$, where $\alpha _{c}\simeq
0.54$, as shown in Fig. \ref{Phase diagram}. The critical exponent is found
to be $\nu \simeq 0.69(2)$ by the linear fit, as shown in Fig. \ref%
{log-log-delta}. It is close to the standard $O(3)$ value of $0.7112(5)$
(e.g. Refs. \onlinecite{Matsumoto,Subir Sachdev,K. Chen}). In comparison to
the result $\alpha _{c}=0.27$ of the nonlinear $\sigma $ model method \cite%
{Takano} and the variational result $\alpha _{c,var}=0.5$ obtained through
Eq. (\ref{trial psi}), the present TRG result is closer to $\alpha _{c}=0.576
$ and $\nu =0.707$ of QMC calculations \cite{F.-J. Jiang}. As expected,
owing to its lower coordinates, the disordered region on the DHC lattice is
wider than that on a square lattice where $\alpha _{c}=0.397$ \cite{Sandro
Wenzel}.

\begin{figure}[tbp]
\includegraphics[angle=0,width=0.8\linewidth]{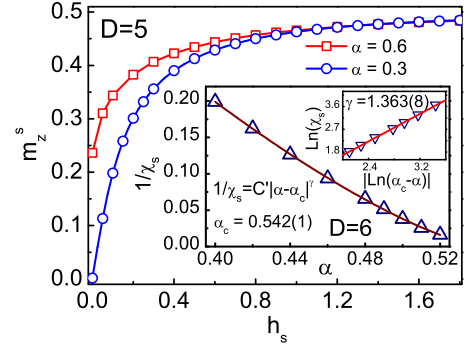}
\caption{(Color online) The staggered magnetization per site $m_z^s$ as a
function of the staggered magnetic field $h_s$ for $h=0$. The inset is $1/%
\protect\chi_s$ against $\protect\alpha$, where the fitting curve is
obtained at a small field $h_s=0.001J$.}
\label{staggered field}
\end{figure}

In the presence of a staggered magnetic field $h_{s}$, the staggered
magnetization per site, $m_{z}^{s}=(1/N)\sum_{i}\epsilon _{i}\langle
S_{i}^{z}\rangle $, as a function of $h_{s}$ for different $\alpha $ is
presented in Fig. \ref{staggered field}, where $h=0$. With increasing $h_{s}$%
, $m_{z}^{s}$ increases monotonously from zero for $\alpha <\alpha _{c}$,
while for $\alpha >\alpha _{c}$, $m_{z}^{s}$ starts from a nonzero value,
implying again that the system has a spontaneous AFLRO for larger $\alpha $.
The inverse of staggered susceptibility $\chi _{s}=\partial
m_{z}^{s}/\partial h_{s}$ as a function of $\alpha $ is presented in the
inset of Fig. \ref{staggered field} for $h=0$ and a very small $h_{s}$. The
critical behavior of the zero-field staggered susceptibility $\chi _{s}$ is
expected to diverge as $|\alpha -\alpha _{c}|^{-\gamma }$ near the critical
point $\alpha _{c}$, where the nonlinear curve fitting gives $\alpha
_{c}\simeq 0.542(1)$ and the log-log plot in the inset shows the critical
exponent $\gamma \simeq 1.363(8)$, which agree with $\alpha _{c}$ obtained
through the spin gap and the $O(3)$ value of $\gamma =1.373(3)$ \cite{K.
Chen}, respectively. According to the obtained critical exponents $\beta $
and $\gamma $, it is seen that the quantum criticality of the present system
falls into the $O(3)$ universality class. Similar calculations indicate that
the transition between the dimer and canted N\'{e}el phases have the same
critical exponents $\beta $ and $\gamma $ as that of the dimer-collinear N%
\'{e}el phase transition at $h=0$. Hence, they also belong to the same $O(3)$
universality class.

\section{Relation to the 2D Boson Hubbard model}

Finally, we would like to mention briefly that the present spin-1/2 system
has a link with the 2D boson Hubbard model. By performing the hard-core
boson mapping\cite{Matsubara} with $a_{i}^{\dag }\rightarrow S_{i}^{+}$, $%
a_{i}\rightarrow S_{i}^{-}$, $n_{i}\rightarrow S_{i}^{z}+1/2$, and making a
rotation of spins by $\pi $ along the $z$ axis on one sublattice, one can
obtain, from Eq. (1) with $h_{s}=0$, the Hamiltonian of hard-core bosons
\begin{eqnarray}
H_{b} &=&\sum_{\langle i,j\rangle _{x}}[-t(a_{i}^{\dag }a_{j}+a_{j}^{\dag
}a_{i})+U(n_{i}-1/2)(n_{j}-1/2)]  \notag \\
&+&\sum_{\langle i,j\rangle _{yz}}[-t^{\prime }(a_{i}^{\dag
}a_{j}+a_{j}^{\dag }a_{i})+U^{\prime }(n_{i}-1/2)(n_{j}-1/2)]  \notag \\
&-&\mu \sum_{i}(n_{i}-1/2),  \label{Hb}
\end{eqnarray}%
where $t=U=J$, $t^{\prime }=U^{\prime }=J^{\prime }$ and $\mu =h$. It is the
boson Hubbard model on the DHC lattice, whose behavior can be well
understood in accordance with the aforementioned corresponding Heisenberg
spin system. With the above mapping, owing to the nonvanishing sublattice
magnetization $|m_{x,A(B)}|$ in $x$-$y$ plane (see the inset of Fig. \ref%
{Magnetization}), the canted N\'{e}el phase in the spin system
corresponds to the boson superfluid phase with off diagonal
long-range order (ODLRO) \cite{Pochung Chen, Stefan Wessel} in the
boson Hubbard model for $t^{\prime }/t=U^{\prime }/U>0.54$ on the
DHC lattice. For $t^{\prime }/t=U^{\prime }/U<0.54$, the disordered
dimer phase in the spin system is mapped onto a liquid state with
neither ODLRO nor DLRO. Therefore, the boson Hubbard model on the
DHC lattice has a QPT from the boson liquid to a superfluid. The
spin polarized phase becomes a Mott insulator phase with one boson
occupying each site in the mapped boson system, and hence, there
also exists a superfluid-Mott insulator transition.

\section{Conclusion}

In conclusion, the spin-$1/2$ HAF model on the DHC lattice is studied by
means of the combined iTEBD and TRG algorithm, where the ground-state phase
diagram is obtained. It is uncovered that there is a second-order QPT from
the disordered dimer phase to the ordered collinear N\'{e}el phase at $%
\alpha _{c}\simeq 0.54$, where the critical exponents $\nu \simeq 0.69(2)$
and $\gamma \simeq 1.363(8)$ are determined. This QPT belongs to the
standard $O(3)$ universality class. In addition, through a boson mapping,
the present HAF system has a link with the boson Hubbard model on the DHC
lattice. The properties of the latter boson Hubbard model can thus be
understood in terms of the present study. We expect that our findings is not
only useful for understanding experimental observations of the
antiferromagnets with DHC lattices, but also is helpful for the
corresponding 2D boson Hubbard model.

\acknowledgements

The authors are grateful to H. C. Jiang, Z. Y. Xie, and T. Xiang for
stimulating discussions. We are also indebted to Z. Y. Chen, Y. T. Hu, Z. C.
Wang, Q. B. Yan, F. Ye, and Q. R. Zheng for useful discussions. This work is
supported in part by the NSFC (Grant Nos. 10625419, 10934008, 90922033), the
MOST of China (Grant No. 2006CB601102) and the Chinese Academy of Sciences.

\end{document}